\documentclass[aps,prc,amsmath,amssymb,amsfonts,reprint,showpacs]{revtex4-1}
\usepackage{graphicx}
\begin{document}
\title{
Multiplicity-dependent saturation momentum in $p$-Pb collisions at 5.02 TeV} 
\author{Takeshi Osada}
\email[]{t-osada@tcu.ac.jp}
\affiliation{
Department of Natural Sciences, Faculty of Science and Engineering\\ 
Tokyo City University, 
Tamazutsumi 1-28-1, Setagaya-ku, Tokyo 158-8557, Japan}
\date{\today}
\begin{abstract}
Semi-inclusive transverse momentum spectra observed in proton-proton and proton-lead nuclear collisions at LHC energies 
obey a geometric scaling with a scaling variable using multiplicity-dependent saturation momentum. 
The saturation momentum extracted from the experimental data is proportional to the 1/6 power of the hadron 
multiplicity in the final state. 
On the other hand, the system's transverse size is proportional to the 1/3 power of the multiplicity, and the saturation 
momentum and the transverse size of the system are strongly correlated with the hadron multiplicity in the final state.   
Since the saturation momentum is proportional to the average transverse momentum of hadrons, 
one predicts average transverse momentum is also proportional to the 1/6 power of the multiplicity, 
which is consistent with experimental results at the LHC energy. 
We found that a nuclear modification factor $R_{\rm pPb}$ calculated by the multiplicity-dependent saturation momentum decreases 
at $p_{\rm T} \lesssim$ 1GeV/$c$ and that our model can partially explain the  $R_{\rm pPb}$'s behavior 
thought to be caused by nuclear shadowing.
On the other hand, Cronin enhancement experimentally observed at $2 \lesssim p_{\rm T} \lesssim $ 6~GeV/$c$ is not reproduced.
However, the experimental result, including the Cronin effect, can be reproduced well by introducing $p_{\rm T}$ 
dependence as a 4$\sim$5\% correction to the multiplicity-dependent saturation momentum. 
We also discuss a relation between the geometric scaling in the semi-inclusive distributions and the string percolation model. 
\end{abstract}
\maketitle
\section{Introduction}\label{sec:01}
The gluon saturation picture \cite{Kovchegov:2012mbw, Gribov:1984tu, Blaizot:1987nc,Kharzeev:2000ph} 
has provided us with many hints for a unified understanding of multi-particle production in which strong interactions 
play a significant role. 
For example, in a Color Glass Condensate (CGC) model \cite{Iancu:2002xk,Blaizot:2004px} 
which is an effective theory to describe saturated gluons with small $x$, 
%as classical color fields radiated by color sources at higher rapidity, 
the saturation scale \cite{Kharzeev:2001gp,Gelis:2010nm} separates 
the classical gluon field into fast frozen color sources and slow dynamical color fields. 
The existence of the intrinsic scale of the transverse momentum $Q_s(x)$ is a crucial underlying 
assumption of the effective theory.
Furthermore, by replacing $p_{\rm T}$ in a Bjorken $x$ of the saturation scale $Q_{s}(x)$ 
with some constant characteristic one, we introduce the energy-dependent 
saturated momentum $Q_{\rm sat}(W)$, which depends only on the collision energy $W$. 
Then, it is a unique scale that governs $p_{\rm T}$ spectra of 
the produced particles, and as a result, geometric scaling \cite{Stasto:2000er,Iancu:2002tr, Iancu:2003jg} 
(GS) is emerges. 
The authors of Ref.\cite{Praszalowicz:2011tc,Praszalowicz:2013fsa,McLerran:2014apa,Praszalowicz:2015dta,Praszalowicz:2015hia} 
confirmed GS certainly holds for inclusive $p_{\rm T}$ spectra of 
high-energy $pp$, $pA$, and $AA$ collisions.  

In our previous work \cite{Osada:2017oxe,Osada:2019oor}, we confirmed that GS also holds even for semi-inclusive distributions. 
In these cases, we introduced a saturation momentum $Q_{\rm sat}(W^*)$ 
that depends on the effective energy $W^*$, which has a one-to-one correspondence with the observed 
multiplicity in the final state instead of the initial colliding energy $W$.
This paper will discuss based on a perspective that the physics of gluon saturation is a fundamental property and should serve as a comprehensive explanation of multi-particle production at different reaction types, energies, and multiplicities.

Recently, a collective motion thought to be characteristic of the hadronic matter produced by 
the collisions of large systems such as $AA$ has been found in high-multiplicity events by 
small systems such as $pp$ and $pA$ collisions \cite{Preghenella:2018moc,Khachatryan:2016txc}. 
Therefore, another hint to the unified understanding of multi-particle production in any type of reaction 
would be seen the similarity in high multiplicity events of $pp$ and $pA$ collisions. 
The multiplicity dependence on the mean transverse momentum in $pp$, $pA$, and $AA$ collisions is 
impressive because their dependence is significantly different for each reaction type, 
especially at high multiplicity \cite{Abelev:2013bla}.  
In particular, theoretical studies need to explain a result that the multiplicity dependence 
on the mean transverse momentum of $p$-Pb collisions is weaker than that of $pp$ for $dn/dy \gtrsim 20$. 

The so-called cold nuclear matter effects observed in $pA$ 
collisions has been investigated by experiments at RHIC \cite{PHENIX:2013fis, PHENIX:2018xqy, PHENIX:2019uuu}
and LHC energies \cite{ALICE:2012mj,ALICE:2014fcr,CMS:2016deq}, 
and theoretical explanations have been added 
to those results \cite{Rezaeian:2008ys,Rezaeian:2009xy,Dumitru:2011wq}. 
An important observable, nuclear modification factor $R_{pA}$, is defined by a ratio of the $p_{\rm T}$  spectrum of $pA$ collisions 
to that of $pp$ collisions, with particular attention paid to the increase in the yield of $p$-Pb collisions 
known as the Cronin effect \cite{Cronin:1974zm,Kharzeev:2003wz,Iancu:2004bx}.
One considers the deviation of the value of $R_{pA}$ from 1 as the nuclear matter effects on particle production, 
making it possible to investigate the multiple scattering effects in the nuclear medium 
including nuclear shadowing \cite{Arneodo:1992wf} and transverse momentum broadening \cite{Accardi:2002ik}. 
One may also extract information on the small $x$ gluon distribution of 
a nucleus in the early stage of collisions \cite{Dumitru:2018iko}. 
The CGC formalism has been successful in explaining these nuclear shadowing, 
and transverse momentum broadening in $pA$ collisions at the LHC \cite{JalilianMarian:2011dt}. 
For collisions of different system sizes, such as $pA$, two saturation momentum scales, i.e., 
$Q_{\rm s}^p$ for proton and $Q_{\rm s}^A$ for nucleus, are introduced into theoretical models. 
However, due to less constrained the initial value of those saturation scales, 
it gives theoretical uncertainties of the nuclear modification factor $R_{\rm pA}$ at LHC \cite{Rezaeian:2012ye}.   
It has also been pointed out that fluctuations in protons' saturation momentum play 
a significant role in the multiplicity distribution of produced particles in $pA$ collisions \cite{McLerran:2015lta}.  

In this paper, the multiplicity-dependent saturation momentum is extracted using the geometric scaling property of the 
semi-inclusive $p_{\rm T}$ spectra in $pp$ and $p$-Pb collisions. 
Furthermore, using the experimental results on the nuclear modification factor in the central rapidity region, 
we further investigate the saturation momentum that governs the multi-particle production process in $p$-Pb collisions.

This article is organized as follows.
We briefly explain the geometric scaling for the semi-inclusive distribution and determine its parameters in Section \ref{sec:semi_incl}. 
Besides, by fitting a universal function of GS to the semi-inclusive $p_{\rm T}$ spectra observed in $pp$ and $p$-Pb collisions 
at LHC energies, 
we determine the multiplicity-dependent saturation momentum $Q_{\rm sat}(W^*)$ and the effective interaction radius 
$R_{\rm T}^*$ as a function of the multiplicity density in the central rapidity region. 
%GS suggests that the mean transverse momentum $\langle p_{\rm T}\rangle$ is proportional 
%to the 1/6 power of the multiplicity because $Q_{\rm sat}$ is proportional to $\langle p_{\rm T}\rangle$.  
Then, we show that the experimental results on the multiplicity dependence of $\langle p_{\rm T}\rangle$ are 
consistent with the GS's conjecture in Sec.\ref{sec:mean_pt}. 
In the Sec.\ref{sec:nucl_mod}, we clarify the role of the saturation momentum in the nuclear modification factor 
$R_{\rm pPb}$ by comparing our model calculations using $Q_{\rm sat}(W^*)$ obtained for $pp$ and $p$-Pb, respectively.  
We close with Sec.\ref{sec:final} containing the summary and some concluding remarks.

\section{GS in semi-inclusive transverse momentum spectra}\label{sec:semi_incl}
Consider transverse momentum spectra of $pp$ or $p$-Pb collisions with colliding (center of mass) energy $W$ 
classified by the multiplicity of the charged hadrons in its final state. 
In the following formulation of our model, except for a determination part of the multiplicity-dependent saturation momentum 
and the universal functions, we follow the theoretical formulation developed in 
Ref. \cite{Osada:2017oxe,Osada:2019oor} for $pp$ collisions.  

For each event multiplicity classes, the semi-inclusive transverse spectra of hadrons 
normalized by a effective crossectional reaction area $S^*_{\rm T}$ 
can be scaled to an universal function 
\begin{subequations}
\begin{eqnarray}
\frac{1}{S^*_{\rm T}}\frac{1}{2\pi p_{\rm T}}\frac{d^2 n_{\rm ch}}{d p_{\rm T}dy}={\cal F}(\tau),\label{eq:semi_incl_GSa}
\end{eqnarray} \label{eq:semi_incl_GS} 
with a scaling variable 
\begin{eqnarray}
\tau^{1/(2+\lambda)} \equiv \frac{p_{\rm T}}{Q_{\rm sat}(W^*)}, 
\label{eq:semi_incl_GSb} 
\end{eqnarray} 
instead of the merely transverse momentum $p_{\rm T}$. 
\end{subequations}
Here, $Q_{\rm sat}(W^*)$ denotes a multiplicity-dependent saturation momentum as a function of the effective energy $W^*$  \cite{Osada:2017oxe,Osada:2019oor}. 
Equation (\ref{eq:semi_incl_GSa}) is originally for the gluon $p_{\rm T}$ distribution 
based on the saturation picture \cite{Kharzeev:2000ph,Kharzeev:2001gp,McLerran:2010uc}. 
We assume that the local parton-duality \cite{Azimov:1984np} 
holds in good approximation, and then hadron spectra observed have the same as 
a gluon distribution but different total multiplicity.  
The factor of the effective area $S_{\rm T}^*$ absorbs the ratio of the partons and hadrons' multiplicity as a constant. 
For an inclusive distribution, the saturation momentum (in literature, it is often referred to as 
an average saturation momentum or an energy-dependent saturation momentum) 
\begin{eqnarray}
  Q_{\rm sat}(W)&=&Q_0 \left( \frac{x_0 W}{Q_0} \right)^{\lambda/(\lambda+2)}, 
  \label{eq:saturation_momentum_scale}
\end{eqnarray} 
is uniquely determined by collision energy $W$ 
with constants $x_0=1.0\times 10^{-3}$, $Q_0=$1.0 GeV/$c$, $\lambda=0.22$ \cite{GolecBiernat:1998js,McLerran:2014apa}.  
In our model \cite{Osada:2019oor}, which deals with GS for the semi-inclusive spectrum, 
we determine $W^*$ and $S_{\rm T}^*$ as fitting parameters to the semi-inclusive spectra 
for each multiplicity fixed by the event class. 
Therefore, $Q_{\rm sat}(W^*)$ has a one-to-one correspondence with the multiplicity 
and regarded as a function of the multiplicity. 
Here, we assume that multiplicity-dependent saturation momentum $Q_{\rm sat}(W^*)$ has 
the same energy dependence as that of Eq. (\ref{eq:saturation_momentum_scale}), 
and it is a saturation momentum in which $W$ in Eq. (\ref{eq:saturation_momentum_scale}) 
is just replaced by $W^*$. 

In our model, $S^*_{\rm T}$ and $W^*$ are fitting parameters, which is equivalent to searching 
for $S^*_{\rm T}$ and $Q_{\rm sat}(W^*)$ directly. 
In fact, there is the following relationship between $W^*$ and $Q_{\rm sat}$: 
\begin{eqnarray}
  W^*=\frac{Q_{\rm sat}}{x_0}\left( \frac{Q_{\rm sat}}{Q_0} \right)^{2/\lambda}. 
\end{eqnarray}
The function ${\cal F}$ in Eq. (\ref{eq:semi_incl_GSa}) is called universal function, and Tsallis type function 
is often used in GS \cite{McLerran:2014apa}:  
\begin{eqnarray}
 {\cal F}(\tau)&=&\left[ 1+(q-1)\frac{\tau^{1/(2+\lambda)}}{\kappa}\right]^{-1/(q-1)},
 \label{eq:universal_function}
\end{eqnarray}
where $q$ is a so-called non-extensive parameter and $\kappa$ is 
a constant parameter which connects $Q_{\rm sat}(W^*)$ as an intermediate energy scale 
and hadronization energy scale, freeze out temperature, for example. 
In previous analyses, we have determined $Q_{\rm sat}(W^*)$ by assuming that a universal function for the inclusive spectra and 
that for the semi-inclusive spectra are the same. 
However, the $p_{\rm T}$ spectra  broadens for the high multiplicity events at 7.0 and 13.0 TeV $pp$ collisions. 
(This tendency already can be seen in Fig. 2 of \cite{Osada:2019oor}.)
Therefore, in this paper, we deal only with $\pi^{\pm}$ spectra to exclude particles with large masses that 
may be sencitive in the large $p_{\rm T}$ region. One can also consider light hadrons such as pions 
to be more suitable for our assumption, such as the saturation picture of gluons and the subsequent 
particle production in the central rapidity. 

\begin{figure}
 \centerline{\includegraphics[width=9.5cm]{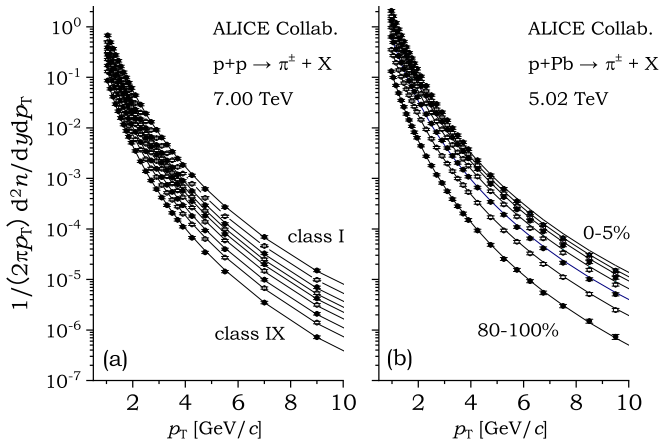}}
 \caption{
 Fit results of $S_{\rm T}^* {\cal F}$ to the semi-inclusive spectra for (a) $pp$ collisions at energy 7.00 TeV \cite{Acharya:2018orn}
 (the multiplicity class is from I~(top) to IX~(bottom),  and the class X is omitted because the multiplicity is too small). 
 (b) The same as (a) but for $p$-Pb collisions at energy 5.02 TeV \cite{Abelev:2013haa,Adam:2016dau} 
 (the multiplicity class is from 0-5\%~(top) to 80-100\%~(bottom)) .
 The value of $q = 1.145$ and $\kappa = 0.1100$ of the universal funcition ${\cal F}$ of Eq. (\ref{eq:universal_function}) is used. 
For the multiplicity of each class, the values of $Q_{\rm sat}(W^*)$ and $S^*_{\rm T}$ are extracted from these fits.
 }\label{fig:1}
 \end{figure}

\begin{figure*}
 \centerline{\includegraphics[width=12.5cm]{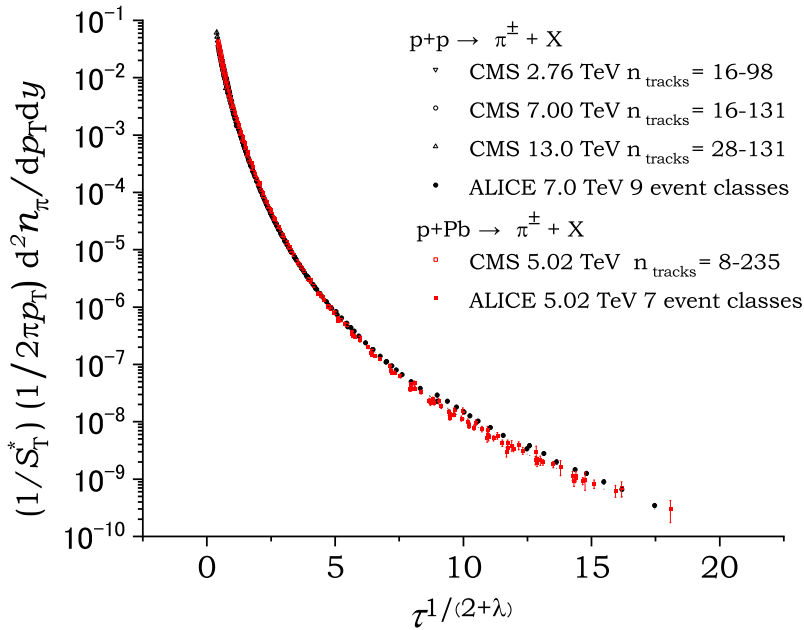}}
 \caption{Geometric scaling of the semi-inclusive $p_{\rm T}$ spectra in $pp$ collisions (black symbols) for multiplicity class 
 with track number $16\leq n_{\rm tracks} \leq 98$ at $\sqrt{s}$=2.76 TeV, 
 $16\leq n_{\rm tracks} \leq 131$ at $\sqrt{s}$=7.00 TeV \cite{Chatrchyan:2012qb}, 
 and  $28\leq n_{\rm tracks} \leq 172$ at $\sqrt{s}$=13.0 TeV \cite{Sirunyan:2017zmn} 
 by CMS Collaboration (the pseudo rapidity window $|\eta|<2.4$).  
The $p_{\rm T}$ spectra in $pp$ collisions for 9 event classes with multiplicity range of 
$3.98 \leq dn_{\rm ch}/dy \leq 20.1$ at 7.00 TeV \cite{Acharya:2018orn} by ALICE Collaboration are also shown. 
For $p$-Pb collisions (red symbols) at 5.02 TeV for multiplicity class with track number $19 \leq n_{\rm tracks} \leq 235$ 
by CMS \cite{Chatrchyan:2013eya} and ALICE Collaboration 7 event classes with multiplicity range of 
$4.4 \leq dn_{\rm ch}/dy \leq 45$ \cite{Abelev:2013haa,Adam:2016dau} are shown.}\label{fig:2}
 \end{figure*}

Since the saturation picture can be universally applied to gluons inside highly relativistic contracted hadronic 
or nuclear matter, GS holds regardless of the collision system and it can be valid in high-energy 
$pp$ and $p$-Pb collisions. 
Therefore, we search for $Q_{\rm sat}(W^*)$ and $S_{\rm T}^*$ in addition to a universal function ${\cal F}(\tau)$ itself, 
which scales semi-inclusive distributions observed in $pp$ and $p$-Pb collisions to the common universal function.
%
%The parameter $q$ and $\kappa$ are determined so that all semi-inclusive distributions correspond to one curve 
%Eq.(\ref{eq:semi_incl_GSa}). 
%At this time, $S^*_{\rm T}$ on the left-hand side in Eq.(\ref{eq:semi_incl_GSa}) 
%and $Q_{\rm sat}(W^*)$ of the scaling variable (\ref{eq:semi_incl_GSb}) are also determined for 
%each semi-inclusive spectra so that they can be regarded as functions of multiplicity. 
%%%%%%%%%%%%%%%%%%%%%%%%%%%%%%%%%%%%%%%%%%%%%%%%%%%%%%%%%%%%%%%%%
%%
\begin{table}
\caption{
Values of $q$ and $\kappa $ in the universal function Eq. (\ref{eq:universal_function}) of geometrical scaling.}\label{tab:1}
\begin{tabular}{cc|ccc}\hline
analysis & observable~ & $q$ & $\kappa$ & $\tau^{1/(2+\lambda)}$ \\ \hline 
$pp$ incl \cite{Osada:2019oor} & charged & ~~1.134~~ & ~~0.1292~~ & $<40$ \\
$pp$ incl & $\pi^{\pm} $ & 1.132 & 0.1111 & $< 2.5$ \\
$pp$ incl & $\pi^{\pm} + K^{\pm} $ & 1.129 & 0.1211 & $<2.5$ \\ \hline 
$pp$, $p$-Pb  semi-incl & $\pi^{\pm}$ & 1.145 & 0.1100  & $<18$ \\ 
\end{tabular} 
\end{table}
%%%%%%%%%%%%%%%%%%%%%%%%%%%%%%%%%%%%%%%%%%%%%%%%%%%%%%%%%%%%%%%%%%
In Fig. \ref{fig:1}, we show two examples of the fit to the semi-inclusive spectra with $q=1.145$, $\kappa=0.1100$ 
for $p+p\to \pi ^ {\rm \pm} + X$ at energy $W=7.00$ TeV and  
for $p+{\rm Pb} \to \pi ^ {\rm \pm}  + X$ at energy $W=5.02$ TeV observed 
by ALICE Collaboration \cite{Acharya:2018orn,Abelev:2013haa}. 
Besides, Fig. \ref{fig:2} shows that 67 semi-inclusive distributions (947 data points), 
including the spectra shown in Fig. \ref{fig:1},  
observed in  $\sqrt{s}=W=2.76, 7.00, 13.0$ TeV $pp$ collisions \cite{
Chatrchyan:2012qb,%CMS 2.76-7.00TeV
Sirunyan:2017zmn,%CMS 13.0TeV
Acharya:2018orn%ALICE 7.00 TeV, 
}, and $W=$5.02 TeV $p$-Pb collisions \cite{
Chatrchyan:2013eya,%CMS pPb 5.02 TeV
Abelev:2013haa,%ALICE pPb 5.02 TeV (PLB728) 
Adam:2016dau%ALICE pPb 5.02 TeV (PLB760) 
} 
almost perfectly scale to the universal function (1a) with $q = 1.145$ and $\kappa = 0.1100$.
% 2.76 CMS 16,28,40,52,63,75,86,98;  => 8 event classes 
% 7.00 CMS 16,28,40,52,63,75,86,98,109,120,131; => 11 event class 
% 13.0 CMS 28,40,51,63,74,85,97,108,119,130,141,151,162,172; => 14 event class
% 7.00 ALICE   => 9 event classes 
% 5.02 CMS 8,19,32,45,58,71,84,96,109,122,135,147,160,173,185,198,210,222,235;  =>19  event class 
%502 ALICE => 7 event class   
% total 68 event classes 1033-67=966 points 
%%%%%%%%%%%%%%%%%%%%%%%%   Insert of Figures  %%%%%%%%%%%%%%%%%%%%%%%%%%%%%%%%%%%%%%%% 
 %%%%%%%%%%%%%%%%%%%%%%%%%%%%%%%%%%%%%%%%%%%%%%%%%%%%%%%%%%%%%%%%%%%%%%%%%%% 
 %%%%%%%%%%%%%%%%%%%%%%%%   Insert of Figures  %%%%%%%%%%%%%%%%%%%%%%%%%%%%%%%%%%%%%%%% 
%%%%%%%%%%%%%%%%%%%%%%%%%%%%%%%%%%%%%%%%%%%%%%%%%%%%%%%%%%%%%%%%%%%%%%%%%%%
As shown in Fig. \ref{fig:2}, we can find suitable $Q_{\rm sat} (W^*)$ and an effective radius of the 
interaction area $R^*_{\rm T}\equiv \sqrt{S^*_{\rm T}/\pi}$ to scale the semi-inclusive $p_{\rm T}$ spectra of both $pp$ and $p$-Pb collisions to the same universal function ${\cal F}(\tau)$. 
Figs.\ref{fig:3} (a) and \ref{fig:4} (a) shows $Q_{\rm sat}(W^*)$ and $R^*_{\rm T}$ extracted 
from the semi-inclusive $p_{\rm T}$ spectra, respectively. 
Both $Q_{\rm sat}(W^*)$ and $R^*_{\rm T}$ are functions of the multiplicity $dn_{\pi}/dy$ 
of the final state pion in the central rapidity region. 
It should be noted that $Q_{\rm sat}(W^*)$ and $R^*_{\rm T}$ are mutually correlated quantities 
because they are subject to the fixed multiplicity constraint of the semi-inclusive event as the following: 
\begin{eqnarray}
  \frac{dn_{\pi}}{dy} 
  = \frac{2\pi\kappa^2 }{(2-q)(3-2q)} ~S^*_{\rm T}Q^2_{\rm sat}(W^*). 
   \label{eq:4_dndy}
\end{eqnarray} 
Considering that $R^*_{\rm T}\propto \left[ dn_{\pi}/dy\right]^{1/3}$ 
approximately holds as well known in the observation of the 
HBT effects \cite{Lisa:2005dd,Aamodt:2011kd,Khachatryan:2011hi},
the saturation momentum 
should be proportional to the 1/6 power of the multiplicity, $Q_{\rm sat}(W^*)\propto \left[ dn_{\pi}/dy\right]^{1/6}$. 
Actually, when $Q_{\rm sat}(W^*)$ and $R^*_{\rm T}$ are plotted by 
$\left[ dn_{\pi}/dy\right]^{1/6}$ and $\left[ dn_{\pi}/dy\right]^{1/3}$, 
respectively, %linear behavior is observed 
one confirms such $dn/dy$ dependence as shown in Figs.\ref{fig:3}~(b) and \ref{fig:4}~(b), 
which is consistent with the simple conjecture expected from GS. 

We discuss somewhat peculiar multiplicity dependence on $Q_{\rm sat}(W^*)$ extracted from $p_{\rm T}$ spectra 
of $p$-Pb collisions at 5.02 TeV observed by the ALICE Collaboration \cite{Abelev:2013haa}. (See, Fig. \ref{fig:3} (a) and (b).) 
It is observed that the saturation momentum extracted from the spectra is 
significantly less multiplicity-dependent than the case of $pp$ collisions.
On the other hand, the extraction of %the multiplicity-dependent saturation momentum 
$Q_{\rm sat}(W^*)$ from spectra observed by 
CMS Collaboration \cite{Chatrchyan:2013eya} with the same collision system and the same energy gives almost 
the same results %for $Q_{\rm sat}(W^*)$ 
as that obtained in $pp$ collisions. 
While ALICE Collaboration has published data on the transverse momentum spectra for $p_{\rm T} <20$ GeV/$c$, 
we used it for $0.6 <p_{\rm T} <3.0$ GeV/$c$ to rule out hadron jet effects 
in the extraction of $Q_{\rm sat}(W^*)$.  
Choosing the maximum $p_{\rm T}=2.0$ GeV/$c$, which is the same as the CMS, 
did not significantly affect results obtained. 
However, the rapidity range is slightly different between two collaborations, where CMS is $|y|<1$, 
whereas ALICE is $0<y<0.5$, and ALICE has  observed pions for the more central rapidity region.
It is still unclear whether the rapidity window of the semi-inclusive $p_{\rm T}$ spectra affects the evaluation of 
$Q_{\rm sat}(W^*)$ and $R^*_{\rm T}$. 

We fitted $Q_{\rm sat}(W^*)$ and $R^*_{\rm T}$ shown in Figs. \ref{fig:3} (a) and Fig. \ref{fig:4} (a) 
by the following fitting formulae of 1/6 and 1/3 power of $dn_{\pi}/dy$, respectively; 
\begin{subequations}
\label{eq:fit_QR}
\begin{eqnarray}
Q_{\rm sat}(W^*)  &=& a_Q+b_Q \left(\frac{dn_{\pi}}{dy} \right)^{\frac{1}{6}}, \label{eq:fit_Qsat}\\
R^*_{\rm T}  &=&a_R+b_R \left(\frac{dn_{\pi}}{dy} \right)^{\frac{1}{3}}. \label{eq:fit_RT}
\end{eqnarray}
\end{subequations}
The values of the coefficients $a_Q$, $b_Q$, $a_R$ and $b_R$ 
%in Eqs.(\ref{eq:fit_Qsat}) and (\ref{eq:fit_RT}) 
fitted to the data of Figs.\ref{fig:3}~(a) and \ref{fig:4}~(a) are shown in Table \ref{tab:2}. 
Here, if the constant terms $a_Q$ and $a_R$ can be ignored, the following relation is derived from Eq. (\ref{eq:4_dndy}); 
\begin{subequations}
\begin{eqnarray}
\sqrt{\frac{(2-q)(3-2q)}{2\pi^2}}=\frac{\kappa~ b_Q b_R}{0.197~\mbox{[GeV$\cdot$fm]}}. 
\label{eq:bQbR}
\end{eqnarray}
As shown in Table \ref{tab:2}, Eq. (\ref{eq:bQbR}) is approximately satisfied by the LHC energies of $pp$ and $p$-Pb collisions.
%%%%%%%%%%%%%%%%%   Insert of Figures  %%%%%%%%%%%%%%%%%%%%%%%%%%%%%%%%%%%%%%%% 
\begin{figure}
\centerline{\includegraphics[width=9.5cm]{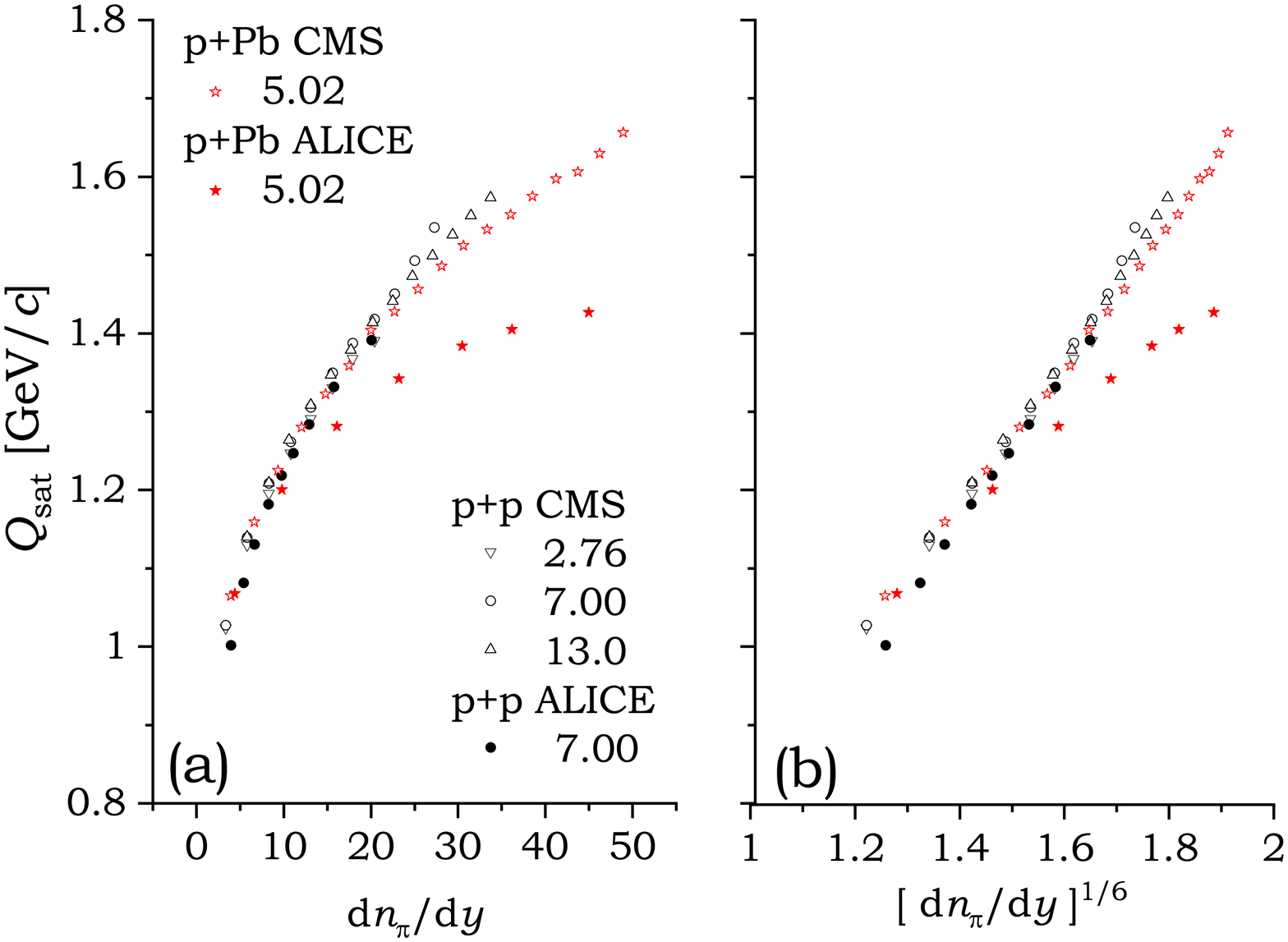}}
 \caption{
(a)~The multiplicity-dependent saturation momentum $Q_{\rm sat}(W^*)$ obtained by 
the fitting Eq. (\ref{eq:semi_incl_GS}) to the semi-inclusive $\pi^{\pm}$ transverse spectra as a function of  $dn_{\pi}/dy$. 
(b)~The same as (a) but as a function of $[dn_{\pi}/dy]^{1/6}$.
}\label{fig:3}
\centerline{\includegraphics[width=9.5cm]{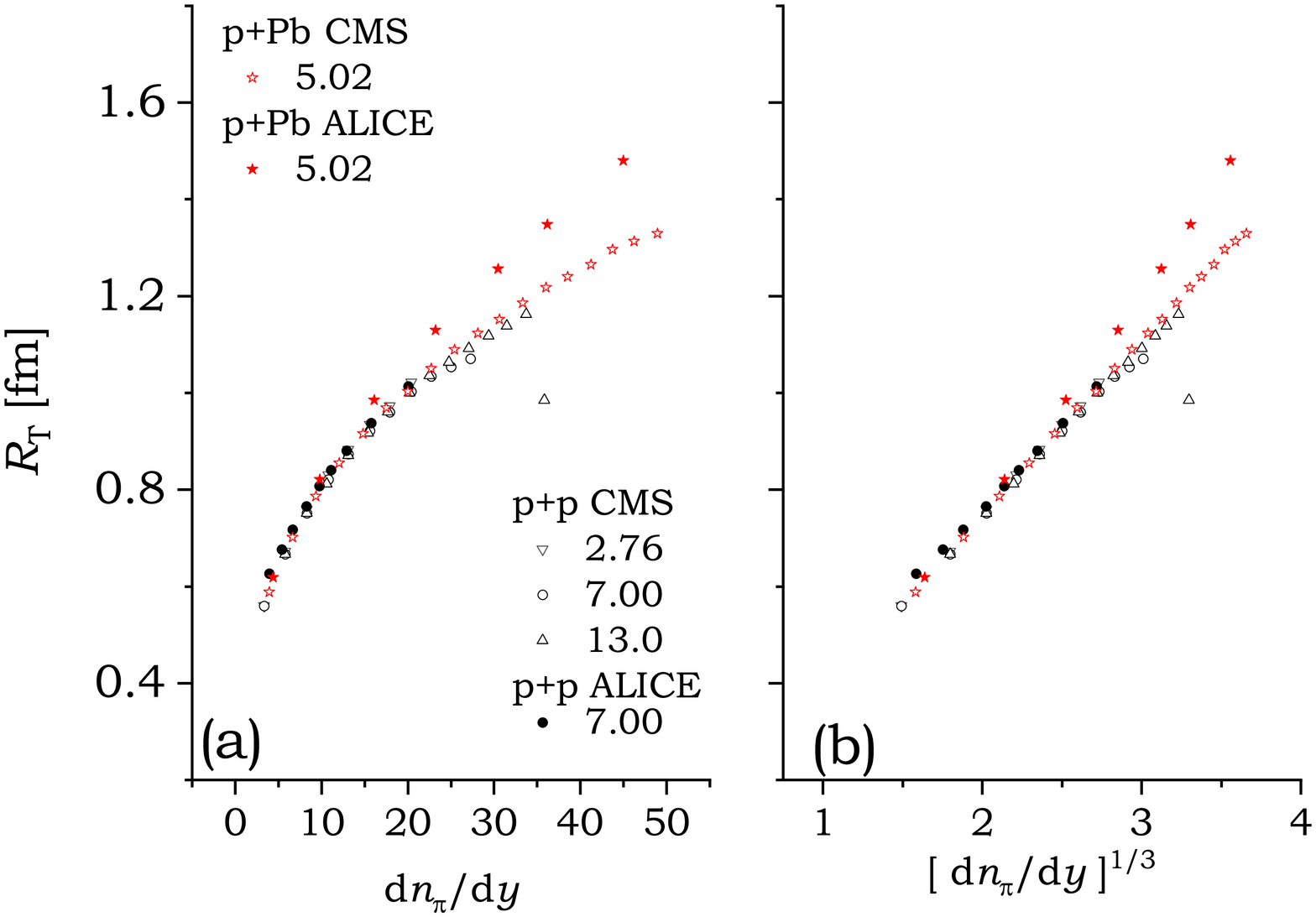}}
 \caption{
(a)~Effective transverse radii $R^*_{\rm T}$ of interaction cross sectional area obtained by the fitting 
Eq. (\ref{eq:semi_incl_GS}) to the semi-inclusive $\pi^{\pm}$ transverse spectra as a function of $dn_{\pi}/dy$
(b)~The same at (a) but as a function of $[dn_{\pi}/dy]^{1/3}$.
} \label{fig:4}
\end{figure}
%%%%%%%%%%%%%%%%%%%%%%%%%%%%%%%%%%%%%%%%%%%%%%%%%%%%%%%%%%%%%%%%%%%%%%%%%%%
%%%%%%%%%%%%%%%%%%%%%%%%%%%%%%%%%%%%%%%%%%%%%%%%%%%%%%%%%%%%%%%%%%%%%%%%%%%
%% 表の挿入
\begin{table*}
\caption{
The values of the parameters used in Eqs. (\ref{eq:fit_Qsat}) and (\ref{eq:fit_RT}) for fitting $Q_{\rm sat}(W^*)$ and $R^*_{\rm T}$ 
extracted from the semi-inclusive $p_{\rm T}$ spectrum, respectively. We also show the right and left hand sides 
of Eq.(\ref{eq:bQbR}) to check the GS conjecture. 
}\label{tab:2}
\begin{tabular}{cc|cc|cc|cc}\hline 
&&\multicolumn{2}{c|}{$~a_Q+b_Q(dn_/dy)^{1/6}$~  } &  \multicolumn{2}{c|}{ $~a_R+b_R(dn/dy)^{1/3}$~ } &\multicolumn{2}{c}{eq.(\ref{eq:bQbR})}\\ \hline \hline 
&& $a_Q$&$b_Q$&$a_R$&$b_R$&l.h.s& r.h.s \\  
$pp$ $\to\pi^{\pm}+X$  && &&& && 
\\ \hline 
2.76 TeV\cite{Chatrchyan:2012qb} &$-1.0<y<1.0$ &~~$-$0.019~~& 0.854 &~~0.006~~ & 0.371 &~~0.175 & ~~0.177\\  
7.00 TeV\cite{Chatrchyan:2012qb} &$-1.0<y<1.0$&~~$-$0.149~~& 0.954 &~~0.051~~ & 0.345 &~~0.175& ~~0.184\\ 
7.00 TeV\cite{Acharya:2018orn}    &$-0.5<y<0.5$&~~$-$0.225~~& 0.985 &~~0.073~~ & 0.345 &~~0.175& ~~0.190 \\ 
13.0 TeV\cite{Sirunyan:2017zmn}  &$-1.0<y<1.0$&~~$-$0.472~~& 1.164 &~~0.156~~ & 0.302 &~~0.175& ~~0.196\\  
$pp$ $\to h^{\pm}+X$  && &&& && \\ \hline
5.02 TeV\cite{Acharya:2019mzb}&$-0.8<\eta<0.8$&~~$-$0.311~~&1.160 &~~0.078~~ &0.296 &~~0.181& ~~0.211\\ 
13.0 TeV\cite{Acharya:2019mzb}&$-0.8<\eta<0.8$&~~$-$0.390~~&1.323&~~0.072~~  &0.246 &~~0.181& ~~0.200\\    
$p$-Pb $\to\pi^{\pm}+X$ &&&&&
\\ \hline  
5.02 TeV\cite{Chatrchyan:2013eya}  &$-1.0<y<1.0$&~~0.078~~& 0.899 &~~0.030~~ & 0.358 &~~0.175&~~0.180\\  
5.02 TeV\cite{Abelev:2013haa}         &$~~0.0<y <0.5$&~~0.315~~& 0.600 &~~$-$0.130~~ & 0.446 &~~0.175&~~0.149\\
\end{tabular} 
\end{table*}
%%%%%%%%%%%%%%%%%%%%%%%%%%%%%%%%%%%%%%%%%%%%%%%%%%%%%%%%%%%%%%%%%%%%%%%%%%%
%%%%%%%%%%%%%%%%%%%%%%%%%%%%%%%%%%%%%%%%%%%%%%%%%%%%%%%%%%%%%%%%%%%%%%%%%%%
At the initial stage of collisions, gluon number density saturates due to their non-linear interactions. 
The inverse of saturation momentum $1/Q_{\rm sat}$ gives a transverse cross-sectional size scale 
where saturated gluons are packed (one may consider it as a color flux tube size)~\cite{Kharzeev:2001gp,Itakura:2008zza,Osada:2017oxe}. 
If we evaluate the tube size scale from inclusive spectra, 
it is determined by solely the collision energy $W$ and does not depend on the event multiplicity.  
For example, when the collision energy is $W= 7.0, 13.0$ TeV, the saturation momentum is $Q_{\rm sat}(W)=$1.213 and 1.289 GeV /$c$, respectively. (These give flux tube size scale 0.162 and 0.153 fm, respectively.)~
However, the saturation momentum obtained from the semi-inclusive spectra is larger than that 
from the inclusive case especially in the high multiplicity event class. 
Therefore, it is considered that the size of the flux tube that appears in the initial stage of collision becomes smaller 
and shrinks slightly as the gluon multiplicity increases.
%Therefore, one may consider that the flux tube size that appears in the early stage becomes smaller in the high multiplicity event class. 
As a result, the tube size appears as a multiplicity dependence such as $1/Q_{\rm sat}(W^*)\propto [dn/dy]^{-1/6}$. 
On the other hand, since the reaction cross-sectional area $S^*_{\rm T}$
becomes large as $S^*_{\rm T}\propto [dn/dy]^{2/3}$, 
one expects the number of flux tubes in the area $S^*_{\rm T}$ to be precisely proportional to $dn/dy$. 
In fact, Fig. \ref{fig:5} (b) shows that the number of color flux tubes extracted from the $pp$ and $p$-Pb 
semi-inclusive events increases linearly from $dn_{\pi}/dy\gtrsim 20$. 
% /////////////////////////////////////////////////////////////////////////////////
Since the size of the color flux tube can be evaluated as $1/Q_{\rm sat}$, 
we get the number of particles produced from a tube per unit rapidity as follows:
\begin{eqnarray}
  \frac{1}{n_{tb}}\frac{dn}{dy}&\sim& \frac{2\pi\kappa^2}{(2-q)(3-2q)}
  =\left[ \frac{0.197 \mbox{[GeV fm]}}{b_Q b_R}\right]^2,\label{eq:bqbr_2}
\end{eqnarray} \label{eq:bqbr_summary}
\end{subequations}
where $n_{tb}=S_{\rm T}^* Q_{\rm sat}^2(W^*)$ is the total number of color flux tubes packed in the effective reaction area. 
Based on the flux tube picture, the above equation near the central rapidity region does not depend on rapidity. 
The change in the slope of 
%transverse momentum spectra as a function of multiplicity 
$Q_{\rm sat}(W^*)$ and $R_{\rm T}^*$ in Fig. \ref{fig:3} and Fig. \ref{fig:4} observed by 
ALICE for $p$-Pb collisions at 5.02 TeV indicates that these $b_Q$ and $b_R$ changes. 
Therefore, it can be considered that the event multiplicity-dependence of both the flux tube size 
and the gluon interaction radius changes at $[dn/dy]^{1/6}\sim 1.6$. 
% /////////////////////////////////////////////////////////////////////////////////
%%%%%%%%%%%%%%%%%%%%%%%%%%%%%%%%%%%%%%%%%%%%%%%%%%%%%%%%%%%%%%%%%%%%%%%%%%%%%
\begin{figure}
 \centerline{\includegraphics[width=8.0cm]{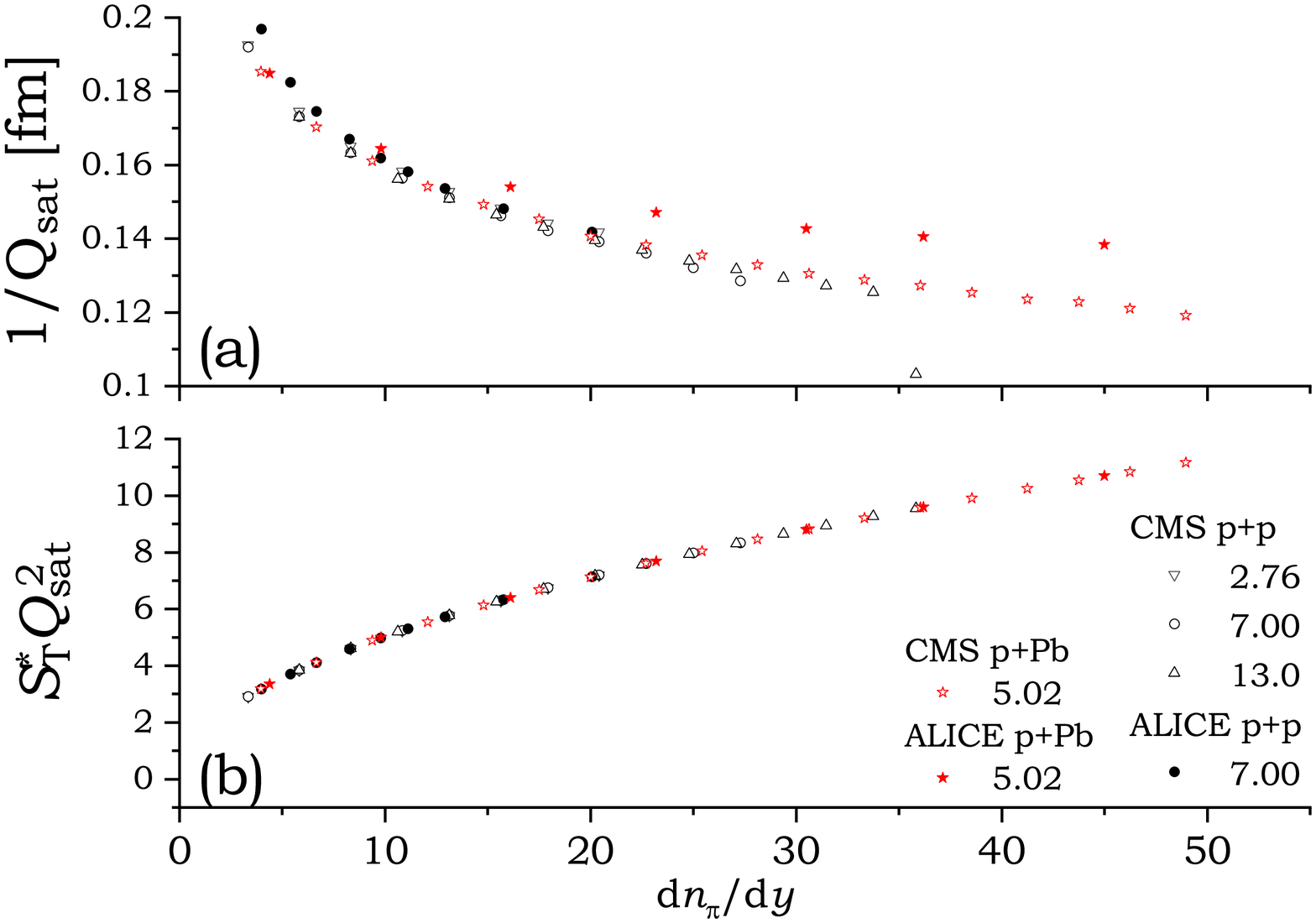}}
 \caption{$dn/dy$ dependence of (a)~$1/Q_{\rm sat}(W^*)$ which can be interpreted as a scale of the radius of color flux tube 
 and (b)~$S_{\rm T}^* Q_{\rm sat}^2(W^*)$ which is the total number of the tubes produced in the interaction area. 
}\label{fig:5}
 \end{figure}
%%%%%%%%%%%%%%%%%%%%%%%%%%%%%%%%%%%%%%%%%%%%%%%%%%%%%%%%%%%%%%%%%%%%%%%%%%%%%
Before closing this section, let us analyze the transverse momentum spectrum obtained by 
the $p$-Pb collision \cite{Abelev:2013haa} in two parts. 
One is the soft part,  
$0.5 \leq p_{\rm T} \leq 1.5~ \mbox{GeV}/c$ (soft $\pi^{\pm}$), 
and the other is the hard part,  
$3.0 \leq p_{\rm T} \leq 19~ \mbox{GeV}/c$ (hard $\pi^{\pm}$). 
By fitting data of each $p_{\rm T}$ window, the saturated momentum $Q_{\rm sat}$  
can be extracted and compared to investigate the relationship 
with the effect of jet quenching \cite{Andres:2012ma}. (Since the measurement 
range of CMS Collaboration is $p_{\rm T}\leq 1.175~ \mbox{GeV}/c$, it is classified 
as soft $\pi^{\pm}$, and the result is the same as a result shown in Fig.\ref{fig:3} (b).)
The multiplicity dependence of $Q_{\rm sat}$ obtained in these two $p_{\rm T}$ windows is shown in Fig.\ref{fig:6}.
The multiplicity dependence of the saturation momentum obtained from soft $\pi^{\pm}$ is not much 
different from results shown in Fig.\ref{fig:3} (b) for low multiplicities. 
On the other hand, for high multiplicity, one observes that $Q_{\rm sat}$ is more clearly proportional 
to the 1/6 power of the multiplicity. 
For hard $\pi^{\pm}$, the multiplicity dependence is the same as the soft $\pi^{\pm}$ case 
up to $[dn/dy]^{1/6}\sim 1.75$, but $Q_{\rm sat}$ is suppressed in a higher multiplicity.
These results may indicate the formation of thin color flux tubes is suppressed by some reason 
in the initial state, or gluons (or pions) with high transverse momentum emitted from thin color flux 
tubes are suppressed due to interactions with hot hadronic mater or cold nuclear matter. 
\begin{figure}
 \centerline{\includegraphics[width=8.0cm]{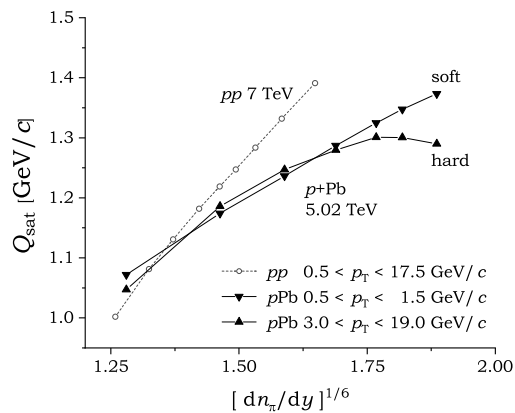}}
 \caption{
 The multiplicity dependence of the saturation momentum $Q_ {\rm sat}$ obtained 
 by fitting Eq. (\ref{eq:semi_incl_GS}) to the semi-inclusive transverse momentum 
 distribution in $p$-Pb collisions for soft $p_{\rm T}$ part ($0.5\leq p_{\rm T} \leq 1.5~{\rm GeV}/c$) and 
 hard $p_{\rm T}$ part ($3.0\leq p_{\rm T} \leq 19~{\rm GeV}/c$). 
 For reference purposes, we also show saturation momentum $Q_ {\rm sat}$ obtained from $pp$ collisions at 7.0~TeV.
 }\label{fig:6}
 \end{figure}
%%%%%%%%%%%%%%%%%%%%%%%%%%%%%%%%%%%%%%%%%%%%%%%%%%%%%%%%%%%%%%%%%%%%%%%%%%%%%
\section{Mean $p_{\rm T}$ in semi-inclusive events}\label{sec:mean_pt}
\begin{figure}
 \centerline{\includegraphics[width=8.5cm]{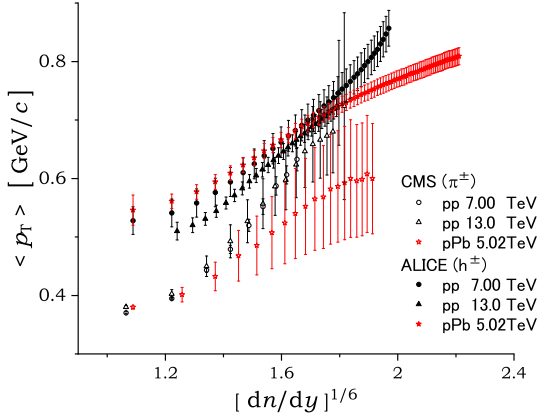}}
 \caption{
Mean transverse momentum $\langle p_{\rm T}\rangle$ of a charged hadron or charged pion as a function of 1/6 power of $dn/dy$. 
The data on the multiplicity $dn/dy$ dependence of  $\langle p_{\rm T}\rangle$ of charged hadrons observed by ALICE Collab.\cite{Abelev:2013bla,Acharya:2019mzb}, 
and the data on charged pions observed by CMS Collab.\cite{Chatrchyan:2012qb,Sirunyan:2017zmn,Chatrchyan:2013eya}}\label{fig:7}
 \end{figure}

The different $dn/dy$ dependence on the average transverse momentum $\langle p_{\rm T}\rangle$ in 
$pp$ and $p$-Pb collisions has been reported in Ref. \cite{Abelev:2013bla}.
%Notably, in $p$-Pb collisions, it is observed that the multiplicity dependence changes at $dn/dy \gtrsim20$. 
In GS model, since the average transverse momentum of the charged hadrons is proportional to the 
multiplicity-dependent saturation momentum $Q_{\rm sat}(W^*)$, one expects $\langle p_{\rm T} \rangle$ 
also proportional to the 1/6 power of the multiplicity fixed for the semi-inclusive events \cite{Osada:2019oor,Osada:2017oxe}: 
\begin{eqnarray}
\langle p_{\rm T} \rangle%(\frac{dn_{\pi^{\pm}}}{dy}) 
= \frac{2\kappa Q_{\rm sat}(W^*)}{4-3q} 
\propto \left( \frac{dn}{dy}\right)^{1/6}. 
\label{eq:6_average_pt}
\end{eqnarray}
Experimental data on the mean transverse momentum of $\pi^{\pm}$ observed by CMS and the mean 
transverse momentum of charged hadron $h^{\pm}$ observed by ALICE are re-plotted in Fig. \ref{fig:7} as 
a function of $dn/dy$ to the 1/6 power. 
As shown in Fig. \ref{fig:7}, the experimental results are substantially proportional to the multiplicity to the 1/6 power. 
It is worth noting that the multiplicity dependence of $\langle p_{\rm T}\rangle$ for $p$-Pb data observed 
by the two experimental groups 
shows similar changes around 
$\left[ dn/dy\right]^{1/6}\sim1.6~(dn/dy \sim 20)$, 
although the absolute value of it differs 
%between the two experimental groups 
due to the difference in acceptance of the measurement. 
These experimental facts suggest that the multiplicity dependence of $Q_{\rm sat}(W^*) $ changes at 
$dn/dy \sim 20$ in central rapidity region. 
Here, ignoring the contribution from $a_Q$ and $a_R$ in Eq. (\ref{eq:fit_QR}) as small 
and using Eqs. (\ref{eq:bQbR}) and (\ref{eq:6_average_pt}),  
we obtain the following for the slope of the graph shown in Fig. \ref{fig:7}; 
\begin{eqnarray}
\frac{\langle p_{\rm T}\rangle}{\left[ dn/dy\right]^{1/6}} &=& \frac{0.197~\mbox{[GeV$\cdot$fm]}}{b_R}\frac{\sqrt{2(2-q)(3q-2)}}{\pi (4-3q)} 
\nonumber \\  
&=& \frac{2\kappa b_Q}{4-3q} \approx  0.3\sim 0.4~[\mbox{GeV}/c]~.\label{eq:slope_pt_1over6}
\end{eqnarray}
The multiplicity dependence of $\langle p_{\rm T}\rangle$ given by Eq. (\ref{eq:slope_pt_1over6}) well explain the behavior of the experimental results. 
Moreover, one may explain the behavior of $\langle p_{\rm T}\rangle$ for $p$-Pb collisions observed by ALICE Collaboration 
for $\left[ dn/dy\right]^{1/6}\gtrsim 1.6$ is due to the change of behavior of the $Q_{\rm sat}(W^*)$. 
Namely, the decrease of $b_Q$ 
and the increase of $b_R$ (See, Table \ref{tab:2}) change the slope of $\langle p_{\rm T}\rangle$ vs. 
$\left[ dn/dy\right]^{1/6}$. 
Note that there is no change in the dependence proportional to the multiplicity's 1/6 power, just a change in the coefficients. 
As pointed out in Sec.\ref{sec:semi_incl}, the saturation 
momentum $Q_{\rm sat}(W^*)$ extracted from the semi-inclusive $p_{\rm T}$ spectra in $p$-Pb collisions 
changes its slope at $\left[ dn/dy\right]^{1/6} \gtrsim1.6$ (See, Fig. \ref{fig:3} (b)). 
Interestingly, both $Q_{\rm sat}(W^*)$ and $\langle p_{\rm T}\rangle$ show a qualitative change 
around the almost same $dn/dy$ in their multiplicity dependence. 
Thus, the saturation momentum that governs the $p_{\rm T}$ spectra 
increases in proportion to the 1/6 power of multiplicity with the same proportional 
coefficient for low multiplicity events in both $pp$ and $p$-Pb collisions. 
However, for high multiplicity events in $p$-Pb collisions,  
$b_Q$ extracted from the ALICE data chages its value at $\left[ dn/dy\right]^{1/6}\approx$1.6. 
Furthermore, as can be seen from Fig. \ref{fig:4}, the coefficient of 
$\left[ dn/dy\right]^{1/3}$ for $R_{\rm T}^*$ also changes at the same multiplicity 
as $Q_{\rm sat}(W^*)$. This is precisely what Eq. (\ref{eq:slope_pt_1over6}) expresses.
On the other hand, for $pp$ collisions, 
there are no indications that the multiplicity dependence of $\langle p_{\rm T}\rangle$ 
changes up to the maximum multiplicity observed. 

This multiplicity dependece change in $\langle p_{\rm T}\rangle$ may be interpreted as follows. 
As can be seen from Fig. \ref{fig:5} (b), the number of flux tubes is proportional to the event multiplicity 
regardless of the reaction and energy. (Approximately four pions are produced per unit rapidity from one tube.) 
In the case of $pp$, the multiplicity increases as the tube's diameter decrease simultaneously as the system's 
reaction size increases. (more tubes are packed in the interaction area.) 
In $p$-Pb, the multiplicity is increased by the same mechanism as the $pp$ collision when the multiplicity is small.  
However, at a certain multiplicity, the flux tube's size becomes difficult to become thin, and instead, 
the reaction region becomes large, so that the multiplicity increases. 
%%%%%%%%%%%%%%%%%%%%%%%%%%%%%%%%%%%%%%%%%%%%%%%%%%%%%%%%%%%%%%%%%%%%%%%%%%%%%%%%%%%%%%%%
%%%%%%%%%%%%%%%%%%%%%%%%%%%%%%%%%%%%%%%%%%%%%%%%%%%%%%%%%%%%%%%%%%%%%%%%%%%%%%%%%%%%%%%%

\section{Nuclear Modification factor}\label{sec:nucl_mod}
In this paper, to introduce the multiplicity dependence of the saturation momentum, 
the originally existing $p_{\rm T}$ dependence of the saturation momentum has been neglected 
by replacing it with a representative $p_{\rm T}$, i.e., $Q_{\rm sat}(W^*)$.
If the nuclear modification factor does not include any final-state interactions, it may be the 
observable where the transverse momentum dependence, which we have been ignoring, 
is most pronounced.
Therefore, we evaluate how much transverse momentum dependence is required for the 
obtained saturated momentum as a correction by comparing it with the 
available experimental data \footnote{The experimental data for nuclear modification factors are obtained using inclusive transverse momentum spectra, including various multiplicities events. Therefore, it is necessary to compare the data on the nuclear correction factors using semi-inclusive events with our model for a more accurate discussion.}.\\

\begin{figure*}
 \includegraphics[width=12.5cm]{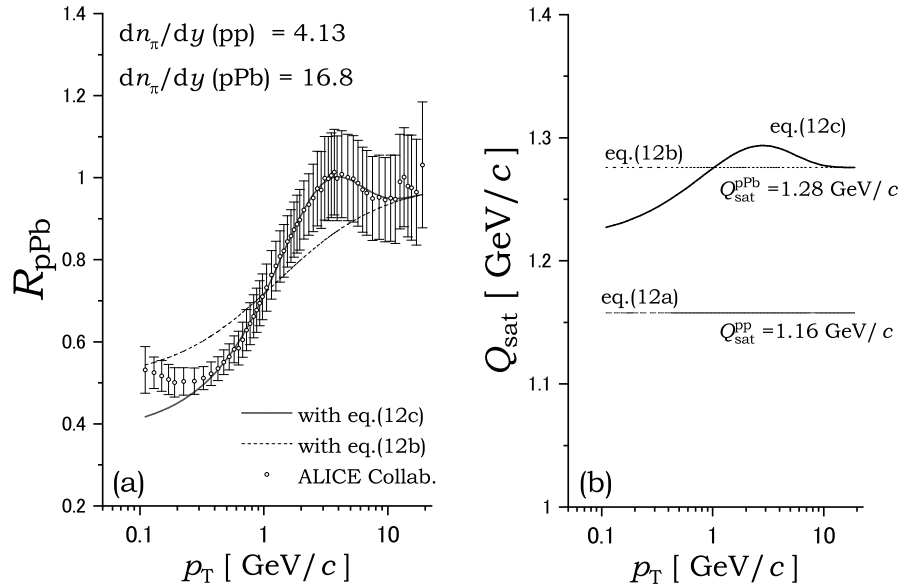}
 \caption{
(a) Comparison of the nuclear modification factor (\ref{eq:R_pA_model}) with 
ALICE data for $p$-Pb collision at 5.02TeV by saturation momentum with Eq(\ref{eq:Qsat_const}) and (\ref{eq:pt_dep_Qsat}) 
shown in the right panel (b). 
} \label{fig:8}
\end{figure*}

We have confirmed in Sec.\ref{sec:semi_incl} that the semi-inclusive $p_{\rm T}$ spectra of 
both $pp$ and $p$-Pb scale to the same universal function. 
The saturation momentum, which plays a central role in the GS, behaves differently from $pp$, 
especially in high-multiplicity events of $p$-Pb collisions. In the case of $p$-Pb collisions, 
nuclear matter may affect the multiplicity-dependent saturation momentum. 
To investigate the nuclear matter effects on the saturation momentum in $p$-Pb collisions, 
the nuclear modification factor experimentally observed is compared with the model calculations. 
The modification factor is a ratio of $p_{\rm T}$ differential yield relative to the $pp$ reference and it is defined by \cite{Adam:2016dau}
\begin{subequations}
\begin{eqnarray}
R^{\rm exp}_{\rm pPb}(p_{\rm T})=\frac{d^2N^{\rm pPb}_{\pi}/d\eta dp_{\rm T}}{\langle T_{\rm pPb}\rangle d^2\sigma^{\rm pp}_{\rm ch}/d\eta dp_{\rm T}},
\label{eq:R_pA_exp}
\end{eqnarray}
where $\langle T_{\rm pPb}\rangle=0.0983$mb${}^{-1}$ \cite{Adam:2016dau} is an average nuclear overlap function. 
In experiments, $R^{\rm exp}_{\rm pPb}$ is defined by the inclusive spectra, but we substitute it 
with the following equation using the semi-inclusive spectra to clarify a role of the saturation momentum: 
\begin{eqnarray}
R_{\rm pPb}(p_{\rm T})=\frac{d^2n^{\rm pPb}_{\pi}/dy dp_{\rm T}}{C~ d^2 n^{\rm pp}_{\rm ch}/dy dp_{\rm T}}. 
\label{eq:R_pA_model}
\end{eqnarray}
\end{subequations} 
Here, $C$ in Eq. (\ref{eq:R_pA_model}) is a constant factor and is related to the experimental data 
of $\langle T_{\rm pPb}\rangle$ and the total inelastic nucleon-nucleon cross section 
$\sigma_{\rm INEL}^{NN}=67.6$ mb \cite{ALICE:2019elz} as follows:
\begin{eqnarray}
   C= \langle T_{\rm pPb}\rangle ~\sigma_{\rm INEL}^{NN}=6.645. 
   \label{eq:constC}
\end{eqnarray}
We apply the multiplicity of semi-inclusive events for $pp$ and $p$-Pb collisions as those of the average multiplicity of inclusive events, respectively: 
i.e., $\langle \frac{dN^{\rm pPb}_{\pi}}{d\eta}\rangle \approx \frac{dn^{\rm pPb}_{\pi}}{dy}=16.81$ \cite{ALICE:2012xs} 
and $\langle \frac{dN^{\rm pp}_{\pi}}{dy}\rangle \approx \frac{dn^{\rm pp}_{\pi}}{dy}=4.13$ \cite{ALICE:2019elz}. 
Therefore, by Eq. (\ref{eq:fit_Qsat}) with values appearing in Table \ref{tab:2}, we have 
\begin{subequations}\label{eq:Qsat_pPb_set}
\begin{eqnarray}
 && Q_{\rm sat}^{pp}(W^*)  =    1.158~\mbox{GeV}/c,  \label{eq:Qsat_const_pp}\\
 && Q_{\rm sat}^{p{\rm Pb}}(W^*) = 1.276~\mbox{GeV}/c. \label{eq:Qsat_const}
\end{eqnarray}
The nuclear modification factor $R_{\rm pPb}$ calculated by Eqs. (\ref{eq:Qsat_const_pp}) and (\ref{eq:Qsat_const}) 
for the multiplicity-dependent saturation momentum of $pp$ and $p$-Pb collisions, respectively,  is shown by the broken line 
in FIG.\ref{fig:8}(a). 
%Using Eqs. (\ref{eq:Qsat_const_pp}) and (\ref{eq:Qsat_const}) for the saturation momentum, respectively, 
We can partially reproduce $R_{\rm pPb}$, such as suppression in the low $p_{\rm T}$ region 
and asymptotic behavior in the high $p_{\rm T}$ region.
However, the simple calculation using Eqs. (\ref{eq:Qsat_const_pp}) and (\ref{eq:Qsat_const}) overestimates 
$R_{\rm pPb}$ in the low $p_{\rm T}$ region compared to the experimental data and 
cannot reproduce the so-called Cronin enhancement. 

Let us introduce $p_{\rm T}$ dependence as a phenomenological side effect on the saturation momentum 
$Q_{\rm sat}(W^*)$, 
which has been regarded as a function of effective energy $W^*$ (or average multiplicity $dn/dy$) only. 
Recall that the saturation momentum $Q_{\rm sat}(W)$ is derived from an intermediate energy scale 
$Q^2_s(x)\equiv Q_0^2(x_0/x)^{\lambda}$ 
given by Bjorken $x=p_{\rm T}/W$. Then $Q_{\rm sat}$ is defined with the solution $p_{\rm T}$ 
satisfying $p_{\rm T}=Q^2_s(p_{\rm T}/W)$. 
Therefore, this is a good approximation for $p_{\rm T}\sim Q_{\rm sat}$ and neglects the weak $p_{\rm T}$ dependence in 
$p_{\rm T} \gg Q_{\rm sat}$ and $p_{\rm T} \ll Q_{\rm sat}$ region, resulting in deviations from the original intermediate energy scale $Q_s(x)$ 
(See Fig. 1 in Ref. \cite{Osada:2019oor}).  
This weak $p_{\rm T}$ dependence may need to be taken into account, especially for observables such as Eq. (\ref{eq:R_pA_model}), 
which is sensitive to the behavior of $p_{\rm T}$. 
Another reason to introduce this effect is to investigate the in unknown detail of gluon recombination effect 
in multi-particle production from experimental data. 
(Of course, it is not possible to distinguish whether recombination occurred before or after the color flux tube formation.) 
We introduce such an effect phenomenologically and investigate whether it 
contributes to explaining $R_{\rm pPb}^{\rm exp}$ obtained in the experiment \cite{Adam:2016dau}.
Thus, instead of Eq. (\ref{eq:Qsat_const}), we introduce modification 
for $Q_{\rm sat}^{p{\rm Pb}}(W^*)$ given by Eq. (\ref{eq:pt_dep_Qsat}) as the following; 
\begin{eqnarray}
   Q_{\rm sat}^{p{\rm Pb}}(W^*) = 
   a_Q+b_Q\left[
   \frac{dn_{\pi}^{p{\rm Pb}}}{dy}
    +\delta
    \right]^{1/6} \!\!\!,
    \label{eq:pt_dep_Qsat}
\end{eqnarray}
where 
\begin{eqnarray}
   \delta=\alpha (\frac{p_{\rm T}-\beta Q^{pp}_{\rm sat}}{Q^{pp}_{\rm sat}})
   \exp\left[\frac{-p_{\rm T}}{\gamma Q^{pp}_{\rm sat}}\right]. 
   \label{eq:pt_dep_Qsat2}
\end{eqnarray}
\label{eq:Qsat_pPb2}
\end{subequations}
We show the results of fitting of Eq. (\ref{eq:R_pA_model}) with 
Eq. (\ref{eq:constC}) and (\ref{eq:Qsat_pPb_set}) to the experimental data $R^{\rm exp}_{p\rm Pb}(p_{\rm T})$ 
by the solid line in Fig. \ref {fig:8} (a),  and the values of the parameters of Eq. (\ref{eq:pt_dep_Qsat2}) in Table \ref {tab:3}. 
As shown in Fig. \ref{fig:8} (b), one can well reproduce the experimental data of the nuclear modification factor 
within 4\% change for saturation momentum $p_{\rm T} \lesssim $ 20 GeV/$c$. 
In particular, the Cronin effect, in which an enhancement peak appears around $p_{\rm T}=2\sim6$ GeV/$c$ 
in $R_{\rm pPb}^{\rm exp}$,  is explained by being about 1\% larger than the multiplicity-dependent saturation momentum of 
$p$-Pb, Eq. (\ref{eq:Qsat_const}). 
Note that the original saturation scale $Q_s(x)$ with fixed collision energy $W$ 
depends on the power of the gluon's transverse momentum; $Q_s(x)\propto p_{\rm T}^{-\lambda /2}$.
For example, when comparing the value of the saturation scale at $p_{\rm T}\approx$ 1 GeV/$c$ 
and 5 GeV/$c$, the value at 5 GeV/$c$ is about 15\% smaller than the value at $p_{\rm T}\approx$ 1 GeV/$c$. 
Hence, the direction of the correction that reintroduces the weak $p_{\rm T}$ 
dependence of $Q_s(x)$ is the opposite direction of the correction required to explain the experimental 
result of $R_{\rm pPb}^{\rm exp}$.

There are two possibilities to explain this variation in the saturation momentum $Q_{\rm sat}(W^*)$. 
One possibility is that the $p_{\rm T}$ dependence in saturation momentum may be explained 
as the effects of interactions such as absorption and emission of gluons after flux tubes decay.  
The other is that such fluctuation may have already existed around the average saturation momentum 
at the time the color flux tube was formed. 
The former suggests that the application of parton-hadron duality requires caution. 
It may be possible to study these two possibilities by similar analyzing a 
nuclear modification factor by using prompt photons \cite{Aaboud:2019tab} in the same way 
as discussed in this article. 
%%%%%%%%%%%%%%%%%%%%%%%%%%%%%%%%%%%%%%%%%%%%%%%%%%%%%%%%%%%%%%%
%% 表の挿入
\begin{table}
\caption{
The values of the parameters used in Eq. (\ref{eq:pt_dep_Qsat2}) giving $R_{p\rm Pb}(p_{\rm T})$ (solid curve) 
in Fig. \ref{fig:8}. }\label{tab:3}
\begin{tabular}{cccccc}\hline\hline 
\quad ${d n^{\rm pPb}_{\pi}}/{dy}$\quad& 
\quad ${d n^{\rm pp}_{\rm ch}}/{d\eta}$\quad   &
\quad  $\alpha$&
\quad  $\beta$&
\quad $\gamma$& 
\quad  $\chi^2$/dof \\ \hline
\quad 16.8 \quad &
\quad 4.13 \quad & 
\quad 6.05\quad & 
\quad 0.886\quad & 
\quad 1.56\quad &
\quad 1.13/46\quad 
\\ \hline 
\end{tabular} 
\end{table}
%%%%%%%%%%%%%%%%%  Figure X  %%%%%%%%%%%%%%%%%%%%%%%%%%%%%%%%%%%%%% 
%%%%%%%%%%%%%%%%%%%%%%%%%%%%%%%%%%%%%%%%%%%%%%%%%%%%%%%%%%%%%%%%%%%%%%%%%%%%%%%%%%%%%%%%

\section{Summary and concluding remarks}\label{sec:final}
Semi-inclusive spectra of $pp$ and $p$-Pb collisions normalized by the 1/3 power of the multiplicity 
scales to the same universal function using a saturation momentum proportional to the 1/6 power 
of the multiplicity. 
The agreement between the experimental results and GS conjecture suggests that 
the saturation momentum, determined by a multiplicity of the final state 
(by assuming local parton hadron duality, it also depends on the gluon's initial state)  
dominates the multi-particle production regardless of the reaction type and collision energy.
The existence of geometric scaling across different reactions such as proton-proton and proton-nucleus collisions 
strongly suggests the gluon saturation mechanism in the early stage of the reaction, which should be a physics 
common to the elementary processes of phenomena on the multi-particle production. 
One of the advantages of GS analysis is that one can derive information on color flux tube formation in the early stage of 
hadronic or nuclear collisions from the multiplicity dependent saturation momentum $Q_{\rm sat}(W^*)$. 
The information may be carried by the coefficients $b_Q$ and $b_R$ in Eqs. (\ref{eq:fit_Qsat}) and (\ref{eq:fit_RT}). 
Moreover, there is a constraint condition between them as Eqs. (\ref{eq:bqbr_summary}) or (\ref{eq:slope_pt_1over6}). 
Based on a color flux tube picture with these equations, 
we pointed out a reason why the multiplicity dependence of mean transverse momentum at high multiplicity in  
$p$-Pb collisions varies compared to $pp$ collisions is a change of the multiplicity dependence of the diameter 
of the color flux tube and the size of the area packed tubes. 
Furthermore, observations of the nuclear modification factor may suggest that the 
multiplicity-dependent saturation momentum needs to introduce small transverse 
momentum dependence. However, the physical origin of this correction for 
the multiplicity-dependent saturation momentum is still unclear.

The string percolation model \cite{DiasdeDeus:2010ejw, Braun:2000hd, Braun:2001us,Braun:1999hv} 
has clear correspondence  with the geometric scaling \cite{Zhang:2014dna} examined in detail in this paper. 
Before closing this last section, let us consider why the multiplicity dependence of 
saturation momentum differs between $pp$ and $p$-Pb collisions using the string percolation model. 
Consider a color string produced in a region $S_{\rm T}$ in $pp$ or $p$-Pb collisions; given the cross-sectional 
area $\sigma_1=\pi r_0^2$ of one string and the string density $\eta=N \sigma_1/S_{\rm T}$ in the area $S_{\rm T}$, 
we have the average number of effective strings $\langle N\rangle$ as the following, (See, Appendix \ref{app:1});
\begin{eqnarray}
\langle N \rangle =S_{\rm T} \frac{(1-e^{-\eta})}{\pi r_0^2 ~F(\eta)}, %\to S_{\rm T}Q_{\rm sat}^2.
\end{eqnarray}
where $F(\eta)$ is a color reduction factor 
\begin{eqnarray}
   F(\eta) = \sqrt{\frac{1-e^{\-\eta}}{\eta}}. 
\end{eqnarray}
By noting that $\langle N\rangle\approx S_{\rm T}Q_{\rm sat}$, 
the corresponding saturation momentum is given by
\begin{eqnarray}
  Q^2_{\rm sat} \propto \left\{ \begin{array}{cc}
       \sqrt{\eta} &\qquad  \mbox{for}\quad\eta \gg  1, \\
        \eta & \qquad \mbox{for} \quad\eta \ll1. \\
        \end{array}
        \right.
 \end{eqnarray}
It is also interesting to note that if $S_{\rm T}$ is proportional to the 2/3 power of multiplicity, 
$\sqrt{\eta}$ (for the case of high multiplicity case) is proportional to the 1/6 power of multiplicity, 
which is consistent with the conclusion 
reached in this paper. 

In the string percolation model, the multiplicity dependence of the saturation momentum 
(the dependence of the inverse of the effective string radius) is explained as a function of the string density.
Therefore, the obtained results that the multiplicity dependence of the saturated momentum 
in $pp$ and $p$-Pb is different for high multiplicity events suggest that the string densities are 
substantially different in $pp$ and $p$-Pb when the multiplicity is sufficiently large.
Our result that the multiplicity dependence of saturated momentum is different for $pp$ 
and $p$-Pb, and smaller for $p$-Pb compared to $pp$, suggests that the color overlap 
is larger for $p$-Pb than for $pp$ when comparing events of the same multiplicity. 
It can also be understood that the transverse reaction area $S_{\rm T}$ with 
the same multiplicity as $pp$ is larger for $p$-Pb, since the multiplicity decreases 
as the string overlap increases. 

The suggestion that there are more overlapping strings in $p$-Pb than in $pp$ may 
be naturally attributed to the effect of nucleons existing around the reaction region 
that are not present in the case of $pp$. 
The color flux tubes (effective strings) that lead to the large multiplicity of final 
states may increase the string density by exciting (or some interactions with) 
the surrounding nuclear matter.
 
The next step in our work is to analyze the semi-inclusive $p_{\rm T}$ spectrum obtained from 
$AA$ collisions to determine a multiplicity dependent saturation momentum. 
Furthermore, the multiplicity dependence of the mean transverse momentum of Pb-Pb collisions is even weaker 
than that of $p$-Pb \cite{Abelev:2013bla}. 
We need to investigate whether multiplicity-dependent saturation momentum extracted from the $AA$ collisions also satisfies the 1/6 power law of 
the multiplicity, and clarify the difference between $p$-A and $AA$ from the viewpoint of a multiplicity-dependent 
saturation momentum. We plan to investigate those issues at some other opportunity. 

\begin{acknowledgements}
This work was supported by JSPS KAKENHI Grant Number 20K03978. 
\end{acknowledgements}

\appendix
\section{String percolation model}\label{app:1}
Suppose that after high-energy $pp$ or $p$-Pb collisions, $N$ 
strings are generated and packed into the cross-sectional area $S$. 
Furthermore, suppose the cross-sectional area of each string is given by $\sigma_1=\pi r_0^2$, 
and the multiplicity generated from one string is $\mu_1$. 
When $n$ strings overlap in an area $S_n$, the color fields in that area are must be summed up like a vector.
Therefore, the multiplicity of hadrons generated from the overlapped $n$ strings is not proportional to $n$ 
but to $\sqrt{n}$. Therefore, as the density of the strings increases, the multiplicity decreases.
Then, the multiplicity produced in the region $S=\sum_n S_n$ is as follows \cite{Braun:1999hv}
\begin{eqnarray}
   \mu = \sum_{n=1}^{N} \frac{\sqrt{n}~S_n}{\sigma_1}\mu_1~.  
\end{eqnarray}
On the other hand, the mean square transverse momentum must be summed like a scaler:
\begin{eqnarray}
  \langle p_{\rm T}^2 \rangle = \frac{N \mu_1}{\mu}\langle p_{\rm T}^2 \rangle_1
  =\frac{N~ \langle p_{\rm T}^2 \rangle_1}{\sum_{n=1}^{N} \sqrt{n}~S_n/\sigma_1}. 
  \label{eq:A2}
\end{eqnarray}
In the limit of $N, S\to\infty$, assuming that the probability of finding the region of $n$ 
strings form a cluster, $p(n)=S_n/S$, obeys the Poisson distribution with the mean value 
\begin{eqnarray}
\eta \equiv N\sigma_1/S=\left( \frac{r_0^2}{R^2}\right) N,
\end{eqnarray}
the multiplicity of the final state is as follows: 
\begin{eqnarray}
  \mu = \frac{S}{\sigma_1} \sum_{n=1}^{\infty} \frac{\sqrt{n}~\eta^n}{n!} e^{-\eta} \mu_1 . 
\end{eqnarray}
Therefore, the color reduction factor $ F(\eta)$, which is defined by the reduction rate 
of multiplicity due to the fusion of strings, is given by 
\begin{eqnarray}
   F(\eta) &\equiv& \frac{\mu}{N\mu_1}= \frac{1}{\eta} \sum_{n=1}^{\infty} \sqrt{n} ~\frac{\eta^ne^{-\eta}}{n!}  \nonumber \\
            &\approx& \sqrt{\frac{1-e^{\-\eta}}{\eta}}. 
\end{eqnarray}
From Eq.(\ref{eq:A2}), we also get the mean squared transverse momentum as follows:
\begin{eqnarray}
   \langle p_{\rm T}^2 \rangle = \frac{\langle p_{\rm T}^2 \rangle_1}{F(\eta)}. 
\end{eqnarray}
Note here that $\sigma_1 F(\eta)$  can be interpreted as the cross-sectional area of the effective string, 
which may be equivalent to the flux tubes in the Glasma picture, formed in the color electric field. 
Hence, the area occupied by the stings $S-S_{0}$ divided by the area of the effective string 
$\sigma_1F(\eta)$ gives the average number of effective strings 
(the average number of color tubes);
\begin{eqnarray}
\langle N \rangle =\frac{S-S_0}{\sigma_1 F(\eta)}=\frac{R^2}{r_0^2}\frac{(1-e^{-\eta})}{F(\eta)}. 
\label{PLB695_eq7}
\end{eqnarray}
Here, noting the correspondence between the effective string and the color flux tube in the 
Glass picture, we have 
\begin{eqnarray}
\langle N \rangle =\pi R^2 \frac{(1-e^{-\eta})}{\pi r_0^2 ~F(\eta)}\to S_{\rm T}Q_{\rm sat}^2.
\label{PLB695_7mod}
\end{eqnarray}
Thus the saturated momentum is related to the string mean density, and we also get
\begin{eqnarray}
  Q^2_{\rm sat}= \frac{\sqrt{\eta~(1-e^{-\eta})}}{\pi r_0^2 ~}. 
 \end{eqnarray}
 Hence, we confirmed from Eq. (\ref{PLB695_7mod}) that the saturation momentum 
 $Q_{\rm sat}$ certainly depends on the multiplicity of the final state hadrons through 
 the string density $\eta$, and the behavior changes as follows depending on whether 
 $\eta$ is large or small: 
 \begin{eqnarray}
  Q^2_{\rm sat} \propto \left\{ \begin{array}{cc}
       \sqrt{\eta} &\qquad  \mbox{for}\quad\eta \gg  1, \\
        \eta & \qquad \mbox{for} \quad\eta \ll1. \\
        \end{array}
        \right.
 \end{eqnarray}
\bibliography{osada1967_2021}
\end{document}